\documentclass{aa}

\usepackage{natbib}
\usepackage{amsmath,amssymb}
\usepackage{hyperref}
\usepackage{bm}
\usepackage{xcolor}
\usepackage{booktabs}
\usepackage{soul}
\usepackage[utf8]{inputenc}

\usepackage{graphicx}
\usepackage{txfonts}

\usepackage{hyphenat}

\newcommand{\msun}{M_\odot}

\newcommand{\un}[1]{\,\mathrm{#1}}

\graphicspath{{imgs}}

\begin{document}
\title{Substructure Detection in Realistic Strong Lensing Systems with Machine Learning}

   \author{Arthur Tsang\thanks{atsang@g.harvard.edu}\inst{1},  Atınç Çağan Şengül\inst{1, 2}, and Cora Dvorkin\inst{1}}

   \institute{Department of Physics, Harvard University,
              Cambridge, MA 02138, USA
              \and
              Department of Physics and Astronomy, University of Pittsburgh, Pittsburgh, PA 15260, USA
    }

   \date{\today}

  \abstract
   {Tens of thousands of galaxy-galaxy strong lensing systems are expected to be discovered by the end of the decade. These will form a vast new dataset that can be used to probe subgalactic dark matter structures through its gravitational effects, which will in turn allow us to study the nature of dark matter at small length scales.}
   {This work shows how we can leverage machine learning to search through the data and identify which systems are most likely to contain dark matter substructure and thus can be studied in greater depth.}
   {We use a UNet, an image segmentation architecture, on a simulated strongly-lensed dataset with realistic sources (COSMOS galaxies), lenses (power-law elliptical profiles with multipoles and external shear), and noise.}
   {Our machine learning algorithm is able to quickly detect most substructure at high image resolution and subhalo concentration. At a false positive rate of 10\%, we are able to identify systems with substructure at a true positive rate of 71\% for a subhalo mass range of $10^{9}\text{-}10^{9.5}\,\msun$. While recent detections are consistent with higher concentrations, we find that our algorithm fails at detecting subhalos with lower concentrations (expected from $\Lambda$CDM simulations).}
   {}

   \keywords{Gravitational lensing: strong -- Methods: data analysis -- Techniques: image processing -- Cosmology: dark matter -- Galaxies: structure
               }

    \titlerunning{Substructure Detection in Realistic Strong Lensing Systems with Machine Learning}
    \authorrunning{Tsang, Şengül, Dvorkin}

   \maketitle
%
\section{Introduction}

Evidence for Lambda Cold Dark Matter ($\Lambda$CDM) usually comes from scales $\gtrsim 10^{11} \,M_\odot$, affecting length scales of order $\gtrsim 1\,\mathrm{Mpc}$ \citep{bullock2017smallscales}. This means that viable alternative theories of dark matter (DM) can differ greatly in their small-length-scale predictions, and that observations at these scales could constrain or distinguish different models. In particular, warm dark matter (WDM) models, such as the gravitino \citep{moroi1993gravitino} or sterile neutrino \citep{boyarsky2009sterile}, predict a certain amount of free streaming, which results in the suppression of structure below a characteristic length scale \citep{bond1983wdm, benson2013wdm,angulo2013wdm}. Likewise, self interacting dark matter (SIDM) can produce similar effects while leaving cold dark matter (CDM) predictions intact at larger scales \citep{tulin2018sidm}.

These smaller-scale structures can correspond to the dwarf galaxies and stellar streams which have been studied in the Local Group \citep{oh2015littlethings, erkal2015gaps, necib2020nyx}, but since smaller gravitationally-bound structures are not conducive to efficient star formation \citep{efstathiou1992cobe, benson2002dwarfsuppression, sawala2016chosenfew}, they are too dim to be observed at greater distances. In addition, typical mass-to-light ratios are very poorly constrained in this regime \citep{wechsler2018galhalo}.

The problems mentioned above are obviated by gravitational lensing, the bending of light from a distant source due to the gravitational field of a lens (see \citet{meneghettibook} for a general reference). This effect lets us probe mass distributions of arbitrarily dim lenses. However, lenses on the scales we want to probe, namely smaller halos and subhalos, produce only subtle deflections and by themselves are degenerate with changes to the source itself. To observe these small deflections, we thus focus on them as perturbations to larger strong lenses, i.e.\ lenses that map the same source onto multiple images. By comparing these images with the assumption of a smooth main lens (i.e. without perturbers), we introduce a consistency requirement that sharpens our sensitivity enough to infer the existence of a perturber.

When a quasar is strongly lensed, one can analyze the relative locations and fluxes of each image to test if they are consistent: a quadruply lensed quasar has a few more degrees of freedom than a reasonable, simple lens model, so it can be used to search for substructure \citep{dalal2003cdmsub}. Many small deviations from the smooth lens model have been found, which can be explained by small-scale DM perturbers (subhalos and line-of-sight interlopers) locally distorting the lens \citep{chiba2002substructure, yonehara2003quasarmesolensing, xu2010quasar,xu2013fluxratio,macleod2013quasarsub, hsueh2019sharp7, birrer2017rxj, gilman2022primordial}. However, with so few extra degrees of freedom, individual subhalo parameters cannot be well constrained, so this method can only be used statistically \citep{dalal2003cdmsub}.

In galaxy-galaxy lenses (the focus of this work), the source is no longer a point source but a galaxy that is warped into so-called Einstein rings or arcs. Like with quasar lensing, the fact that the background galaxy is lensed multiple times adds a degree of redundancy, from which we can measure an integrated line-of-sight shear \citep{birrer17, hogg2023concept}, or as we shall discuss further, identify smaller lensing perturbations, without a priori knowledge of the unlensed source galaxy. These perturbations can be due to both subhalos and line-of-sight interlopers, which create similar but distinguishable effects \citep{li_los, Despali_los, Sengul_los, he2022}. For simplicity, this paper will focus only on subhalos.

The technique of exploiting the consistency constraint between the multiple distorted images works well enough that individual subhalos and line-of-sight halos have already been detected this way \citep{vegetti2010sdss,vegetti2012jvas, hezaveh2016sdp, caganjvas, nightingale2022scanhst}.
Even a lack of a detection, the most common case for the lensing systems studied so far, can be used to put upper bounds on possible amounts of substructure \citep{vegetti2014nondetect, ritondale2019nondetect, nightingale2022scanhst}.

Furthermore, it has been shown that these perturbations can be analyzed statistically as the combined effect of many small subhalos, which itself may provide valuable information on the subhalo mass function and their effective density slopes \citep{hezaveh2016powerspec,diazrivero2018ps,diazrivero2018ethos,zhang2022slopes,wagner2023endtoend}. 

There are currently on the order of a hundred known galaxy-galaxy lensing systems, mostly from SLACS \citep{bolton2008slacs}, but by 2030, we are expected to have discovered tens to hundreds of thousands, from data from the Dark Energy Survey (DES), the Vera Rubin Observatory (VRO), and Euclid \citep{serjeant2014lakhlenses, collett2015forecast, des2016,smith2021vro, euclid2022}. In the context of these new upcoming discoveries, traditional methods of detecting these perturbers are too computationally expensive. To solve the inverse problem, where neither the lens mass nor source light distributions are known, one must sample a high-dimensional parameter space, which takes on the order of days to weeks for a single lens. In order to properly extract all the small-scale structure information contained in this data, we will need new, faster methods of detection.

Since most lensing systems studied so far have resulted in non-detections, it would be useful to quickly flag those likely to contain an individually detectable perturber. This is especially relevant because once we can find perturbers efficiently, we can apply more computationally expensive fitting procedures  to determine the power-law slope of each profile, which can then be used to distinguish between different DM models \citep{cagan2022slope}. Compared to traditional sampling methods, a machine learning model, trained once, can evaluate a large dataset in negligible time, hence the great potential for machine learning to accelerate these searches. In particular, \citet{coogan2020target} showed that machine learning with targeted training sets, i.e.\ custom training sets generated to resemble each lens, can be used to detect and locate individual subhalos. Various forms of machine-learning-based analysis and inference have been proposed by \citet{hezaveh2017cnn}, \citet{biggio2022continuousneuralfields}, \citet{coogan2022tmnre}, \citet{adam2023rim}, \citet{veloso2023lensformer}, \citet{cheeramvelil2023neurips}, and others.

We build off of previous work on this problem using non-targeted machine learning \citep{diazrivero20,ostdiek22apj, ostdiek22letter}, but under more realistic conditions, and we show how this can be used to identify a significant percentage of the galaxy-galaxy lensing systems that contain individually detectable perturbers.

\section{Methods}
\subsection{Simulated lensing systems}

We generate our simulated dataset using \texttt{lenstronomy} \citep{lenstronomy}.
We model the main lens analytically as a power-law elliptical potential (PEP), starting with the assumption that the density $\rho(r) \propto r^{-\gamma}$ and stretching one of the axes to give
\begin{equation}
    \psi(p) = \frac{2E^2}{\eta^2}\left( \frac{p^2}{E^2} \right)^{\eta/2},
\end{equation}
where $\psi$ is the 2D gravitational potential, $p^2 = x_1^2 + x_2^2/q^2$ ($x_1$ and $x_2$ are aligned with the major and minor axes and $q$ is the axis ratio), $\eta = -\gamma + 3$, and $E$ is a normalization factor \citep{barkana1998fast}, which we compute following \texttt{lenstronomy} 
\citep{lenstronomy} as
\begin{equation}
    E = \frac{\theta_E}{((3-\gamma)/2)^{1/(1-\gamma)} \sqrt{q}} ,
\end{equation}
where $\theta_E$ is the Einstein radius in angular units.
In addition to this main lens, we add an external shear, as well as 3rd and 4th order multipoles (the 2nd multipole is equivalent to an ellipticity). Realistic sources are much less amenable to analytic modeling, since while for the lens we are mostly sensitive to the mass distribution in a narrow region of the halo far from the center, the source model must capture the entire stellar light distribution, which is often irregular and may contain spiral arms. For sources, we use rotated and re-centered galaxy images taken from the COSMOS survey \citep{cosmos}, which we interface through the software \texttt{paltas} \citep{wagner2022paltas}, using their default train-test split.  We set a faintest apparent AB magnitude for the COSMOS sources of 20. We randomly vary our lens and source parameters within the ranges shown in Table~\ref{tab:data_params}.

In order for the training process to focus on how to find subhalos while still teaching our machine learning model that an image need not contain one, we add a subhalo to $90\%$ of our images, whose mass varies between $10^8$ and $10^{11}\un\msun$. The most physically relevant subhalo masses for our method are $\lesssim 10^{9.5}\un\msun$, since at larger masses the subhalos become directly visible \citep{sawala2016chosenfew, benitezllambay2020}, but we found that including dark, larger-mass subhalos in the training set still improves performance when tested on lower masses. For consistency with the literature, we use the $M_{200}$ mass, defined as the mass within a radius, $R_{200}$, such that the mean density within this radius is 200 times the critical density of the Universe.
We model our subhalo profile as a truncated NFW,
\begin{equation}
    \rho(r) = \frac{M_0}{4\pi r (r + r_s)^2}\frac{r_t^2}{r^2+r_t^2},
\end{equation}
where $r_s$ is the scale radius, $r_t$ the truncation radius, and $M_0$ is a normalization proportional to the total mass \citep{baltz2009analyticpot}. We set $r_s$ to give a fiducial concentration of $c \equiv R_{200} / r_s = 60$:
\begin{equation}
    r_s = \frac{R_{200}}{c} = \frac{1}{c}\left( \frac{3M_{200}}{4\cdot 200 \pi \rho_{crit}} \right)^{1/3},
    \label{eq:rs200}
\end{equation}
which is in line with recent analyses of observed systems \citep{minor21unexpected, caganjvas, zhang2024hst}. We note that this is in tension with $\Lambda$CDM simulation predictions, so we also test a concentration favored by simulations, $c=15$ \citep{springel2008aquarius}.
We use a fixed ratio for truncation radius to scale radius, $\tau \equiv r_t/r_s = 20$ \citep{baltz2009analyticpot}.

We do not include lens light since it can be subtracted out as a preprocessing step, and we leave a quantification of its effect for future work. Note that improper lens-light subtraction can result in false detections \citep{nightingale2022scanhst}.

The location of the subhalo is chosen uniformly within the area that is at least $20\%$ as bright as the brightest pixel. If the subhalo is located in a part of the image that is too dim, any signal would be overshadowed by noise and we would be unable to detect it regardless of the computational method.

For the sake of simplicity, and since at most one individual perturber has been found in any system analyzed so far, we only include at most one subhalo in each image.

We simulate telescope parameters using \texttt{lenstronomy}'s settings for the Hubble Space Telescope, with the wavelength band WFC3\_F160W and point spread function (PSF) type ``GAUSSIAN.'' We vary the resolution, from 80 mas (the actual (drizzled) resolution of Hubble) to 20 mas and 10 mas, which should be achievable by the upcoming generation of Extremely Large Telescopes (ELTs) \citep{davies2018micado}. We vary the PSF as a function of the resolution: 80 mas for the lower-resolution images to match Hubble and 10 mas for the higher two resolutions to match Extremely Large Telescope projections. We set the noise to the level expected from $10$ exposures corresponding to 90 minutes each. We set the image size to a $6.4$ by $6.4$ arcsec square for all images, varying the number of pixels as appropriate. This size was chosen based on the size of our lensed systems.

\begin{table*}[h]
    \centering
    \begin{tabular}{l c c}
        \toprule
        Parameter && Value or Range \\
        \midrule
        Lens Einstein radius & $\theta_\textit{lens}$ & [0.8, 1.2] arcsec \\
        Lens power-law slope & $\gamma_\textit{lens}$ & [1.5, 2.5]\\
        Lens minor/major axis ratio &  $q_\textit{lens}$ &  [0.5, 1]\\
        Lens orientation & $\phi_\textit{lens}$ & [$-\frac{\pi}{2}$, $\frac{\pi}{2}$]\\
        Lens center & $x_\textit{lens}$, $y_\textit{lens}$ & [$-0.2$, 0.2] arcsec\\
        Lens shear & $\gamma_1$, $\gamma_2$ & [$-0.1$, 0.1] \\
        Lens multipole magnitudes & $a_3$, $a_4$ & [$-0.02$, 0.02] \\
        Lens multipole orientations & $\phi_3$ & [$-\frac{\pi}{3}$, $\frac{\pi}{3}$]\\
         & $\phi_4$ & [$-\frac{\pi}{4}$, $\frac{\pi}{4}$] \\
        \midrule
        Source orientation & $\phi_\textit{source}$ & [$-\pi$, $\pi$]\\
        Source center & $x_{source}$, $y_{source}$ & [$-0.2$, 0.2] arcsec \\
        \midrule
        Subhalo $M_{200}$ mass & $m_\textit{sub}$ & [$10^8$, $10^{11}$] $\msun$\\
        Subhalo concentration & $c_\textit{sub}$ & 15, 60 \\
        Subhalo truncation/scale radius ratio & $\tau_\textit{int}$ & 20\\
        \midrule
        Lens redshift & $z_\textit{lens}$ & 0.5\\
        Source redshift & $z_\textit{source}$ & 1.0 \\
        \midrule
        Image pixel width & & 80, 20, 10 mas\\
        Image angular width & & 6.4 arcsec\\
        \bottomrule
    \end{tabular}
    \caption{Parameter ranges and values used to generate simulated data. Parameters with ranges are taken from a uniform distribution, except subhalo mass, which is log uniform. Subhalo concentration and image pixel width are held constant on each particular dataset.}
    \label{tab:data_params}
\end{table*}

\subsection{Machine learning}
\label{subsec:ml}

\begin{figure}
    \centering
    \includegraphics[width=\columnwidth]{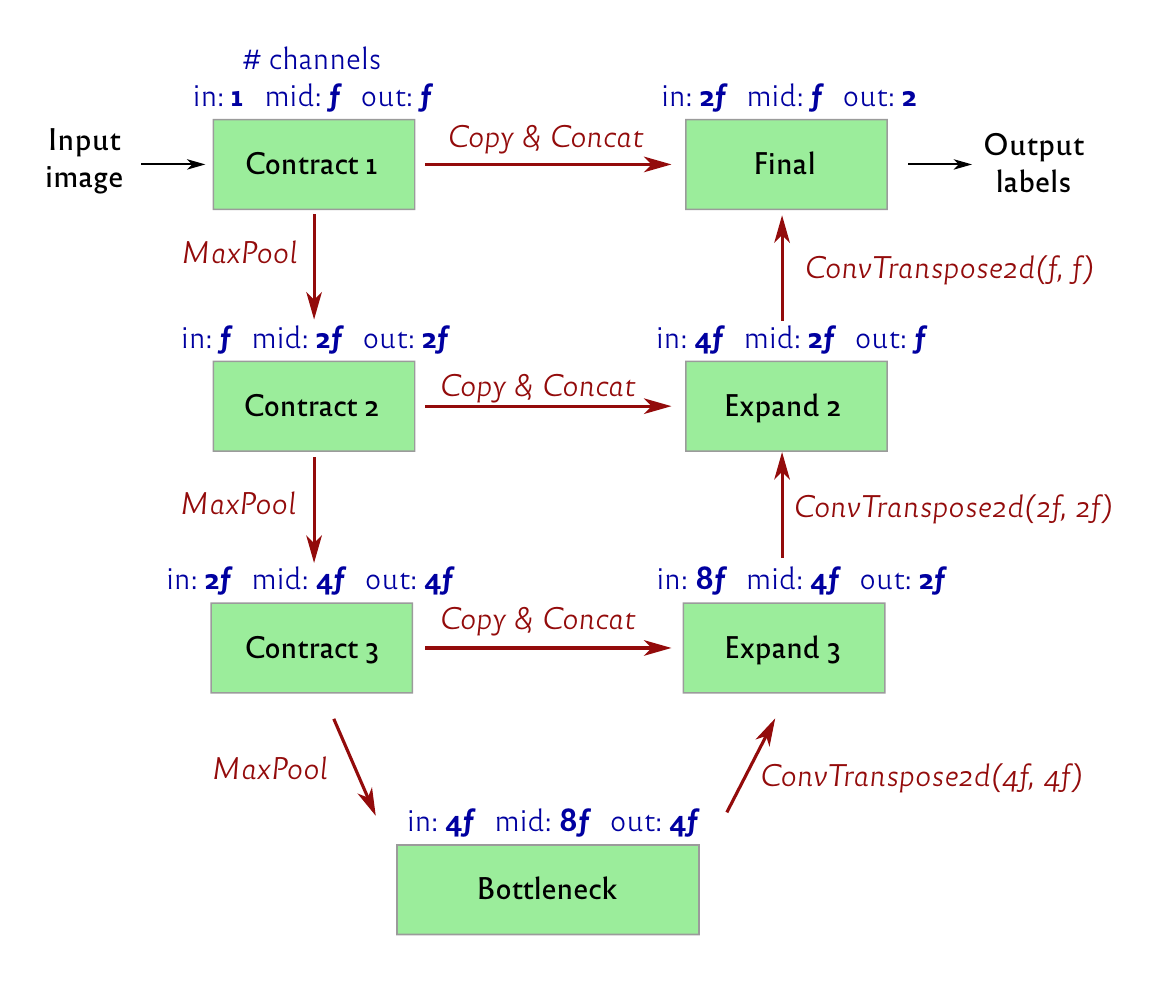}
    \caption{UNet diagram showing the U-shaped architecture composed of several blocks and the operations connecting them. In our case, the input image is a square image of the lensed system, and the output labels are probabilities predicting whether each pixel belongs either to the class \emph{subhalo} or \emph{no-subhalo}. Every block transforms an input tensor with \emph{in} channels into an output tensor with \emph{out} channels (converting it to a tensor with \emph{mid} channels as an intermediate step): these channel numbers are shown above each block, where $f$ is a hyperparameter we set to $32$. See Figure~\ref{fig:unet_blocks} for a more detailed view of the inside of each block. Blocks keep the height and width of their tensors constant, so the MaxPool and ConvTranspose2d operations are used respectively to contract and expand both height and width by a factor of $2$. The Copy \& Concat operation copies the output of the relevant Contract block and concatenates it to the result of the relevant ConvTranspose2d operation, resulting in the input to the Expand block.}
    \label{fig:unet_diagram}
\end{figure}

\begin{figure}
    \centering
    \includegraphics[width=\columnwidth]{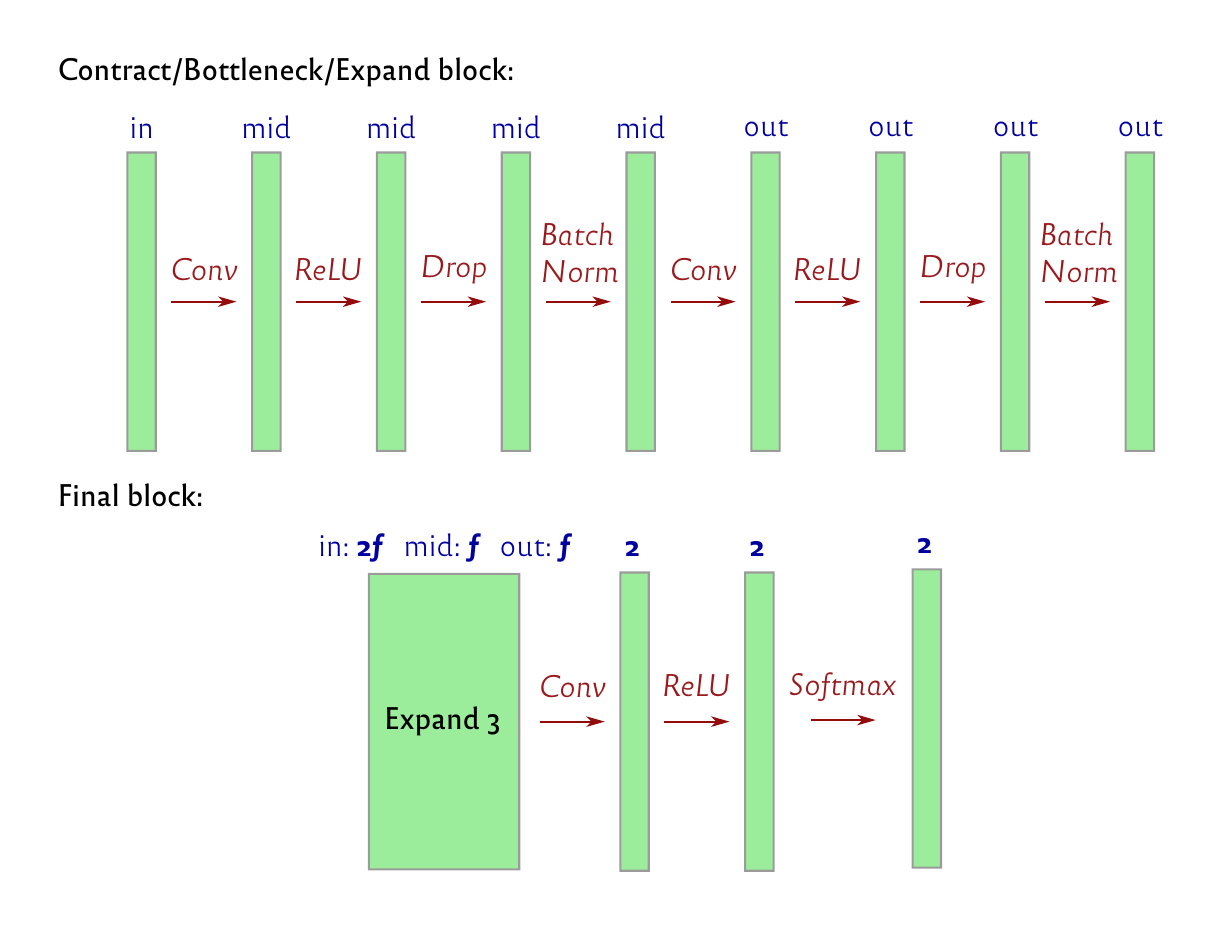}
    \caption{Illustration of the transformations that comprise each block of the UNet. \emph{Top:} A typical block. Each thin bar represents a tensor, with the number of channels written above (\emph{in}, \emph{mid}, or \emph{out}, corresponding to the label in Figure~\ref{fig:unet_diagram}). The arrows represent each sequential operation applied to the tensors, labeled as a convolution, a ReLU nonlinearity, a dropout layer, or a batch normalization. \emph{Bottom:} The final block consists of a typical block followed by several more operations, ending with a final softmax which rescales the probabilities on each pixel to add up to one.}
    \label{fig:unet_blocks}
\end{figure}

Building off of work by \citet{ostdiek22apj}, we again apply a UNet architecture \citep{ronneberger2015unet} to find subhalos using image segmentation. That is, we train a system to label each pixel of the input image according to a predefined set of classes. Like with most machine learning classification tasks, these labels are probabilistic, meaning each pixel's prediction is not given as a particular class, but rather as a probability for belonging to each class. In our case, the input is an image of the strong lensing system (a simulated observation), and the output is a labeling of each pixel with a probability of belonging to one of two classes: \emph{subhalo} or \emph{no-subhalo}. We define the true class of a pixel to be \emph{subhalo} if it is within a certain distance of the center of a subhalo, and \emph{no-subhalo} otherwise; this distance is 2 pixels at lower resolutions (80 mas or 20 mas) and 4 pixels at high resolution (10 mas). We found these distances to work well for a wide range of subhalo masses.

To give a brief sense of the operations used within the UNet, the first is the regular convolution, which involves multiplying the values of an input tensor by a sliding window to produce an output tensor. We refer to the height and width of this window as the kernel size. The Rectified Linear Unit (ReLU) is a simple nonlinear function applied to each tensor element, defined as $f(x) = \max(0, x)$, where $x$ is the input tensor element. The dropout layer, active only during training, randomly zeroes out particular elements of its input with a probability called the dropout rate, in order to reduce overfitting. We train on images not one at a time, but in batches: the batch normalization step re-centers the mean and standard deviation using the values of all the analogous tensors within the given batch. When testing, this layer remembers the mean and standard deviations from training, so images can be evaluated independently. MaxPool is a step that groups an input array into $2\times 2$ squares and reduces each square to the maximum value within it. The tranposed convolution (ConvTranspose2d) has an almost opposite effect, converting each input array element into a $2\times 2$ square.

The UNet architecture applies the convolution operation at different scales, shown as the different horizontal levels in the diagram in Figure~\ref{fig:unet_diagram} (convolutions and other operations are contained within each block as shown in Figure~\ref{fig:unet_blocks}). The intermediate values for a given 2D input image are stored as 3D tensors, or equivalently as a set of 2D images stacked on top of each other, where each 2D image is called a \textit{channel}. This is useful for storing more information about each pixel at each intermediate step. As the diagram in Figure~\ref{fig:unet_diagram} shows, information generally flows in a U-shape. Every level down represents a contraction of the height and width by a factor of 2, rendering the level sensitive to larger-scale features than the level above. An $80\times80$ input image is eventually downsampled to $10\times10$ by the time it reaches the bottleneck. The bottleneck processes the most global and high-level information before it is upsampled and converted into a full-resolution segmentation. As the information flows back up the U, we supplement it by copying and concatenating the last intermediate state on that level (namely the result of the block on the left side of the U), in order to better retain smaller-scale information.

The final layer of the UNet outputs a probabilistic prediction, a tensor with dimensions $(N_\textit{height},\, N_\textit{width},\, N_\textit{class})$, where $N_\textit{class} = 2$ and $N_\textit{height} = N_\textit{width}$ is either 80, 320, or 640, depending on the input resolution, chosen such that the total width of the image is $6.4 \un{arcsec}$. As this is a tensor of predicted probabilities, the values along the last axis sum to $1$, as ensured by the final softmax operation. We train the UNet to minimize the cross-entropy loss,
\begin{equation}
  \mathcal{L}(\theta) = -\sum_{i = 1}^{N_\textit{height}} \sum_{j=1}^{N_\textit{width}} \sum_{k=1}^{N_\textit{class}} q(y_{ij} = k) \log p(y_{ij} = k| \theta) ,
\end{equation}
where $\theta$ represents the UNet parameters, $y$ is a potential class label for each pixel (representing either \emph{subhalo} or \emph{no-subhalo}), $q$ is the ``true'' probability (either $0$ or $1$), and $p$ is the UNet-predicted probability.

To train the UNet with this loss, we use the Adam optimizer \citep{adamoptimizer} with the learning rate scheduler \texttt{ReduceLROnPlateau} available in \texttt{torch} \citep{torch}, with a patience of 5 and a minimum learning rate of $10^{-6}$. After varying hyperparameters, testing one at a time around a well-performing fiducial model, we found the best model had a convolutional kernel size of 3, a batch size of 16, $f = 32$ channels (see Figure \ref{fig:unet_diagram}), $5\times 10^5$ training images, and a 10\% dropout rate.

\section{Results}

\begin{figure*}
    \centering
    \includegraphics[width=.75\textwidth]{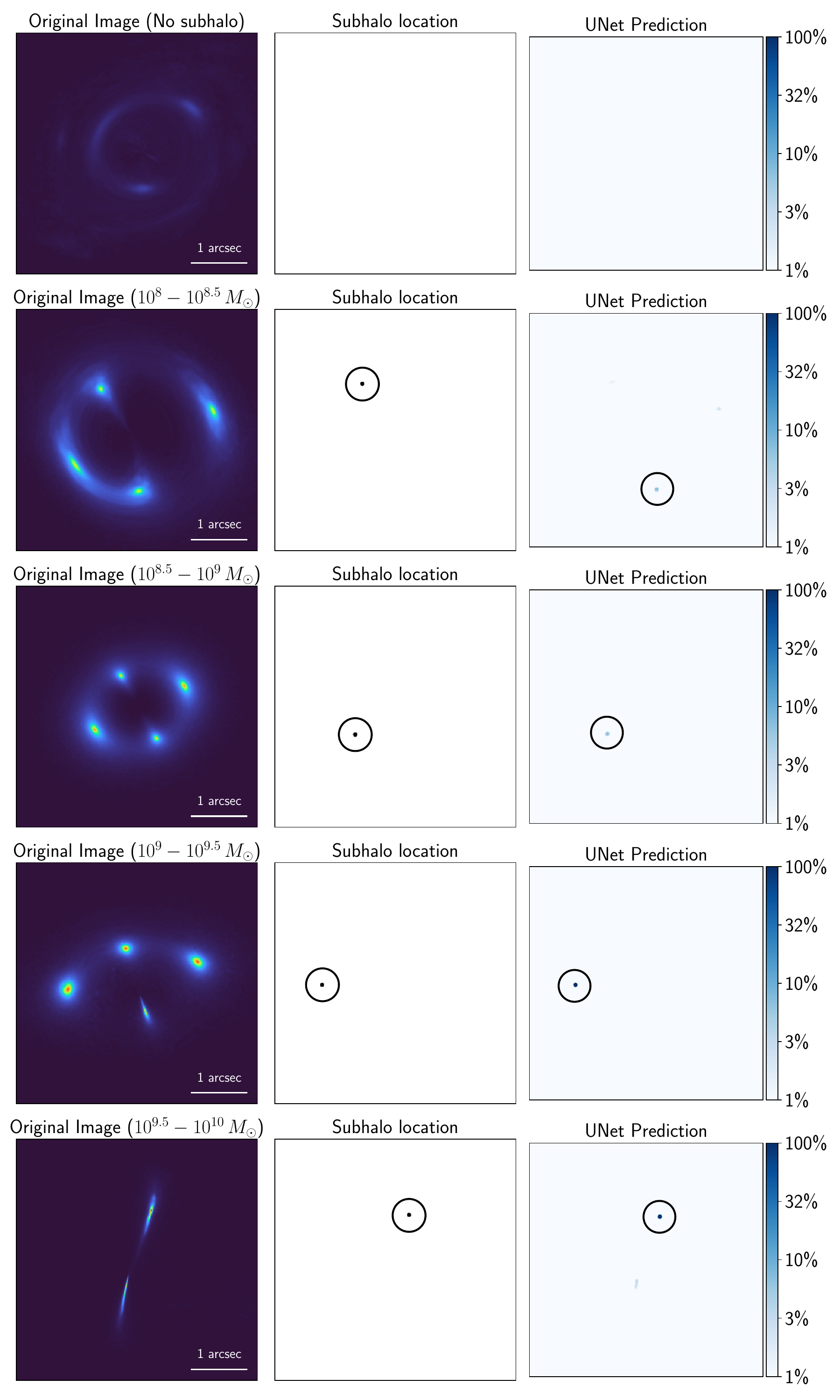}
    \caption{ Characteristic examples on test-set lensing systems. The first column shows the original $640\times 640$ pixels high-resolution image, cropped to $440\times 440$ pixels for visualization purposes. The second column shows the ground truth subhalo location (circled). The third column shows the UNet's predicted probability that each pixel contains a subhalo (predicted position circled). Probabilities are on a log scale to better show the low-probability predictions ($5\text{-}10\%$) on the second and third rows. From top to bottom, we see that as the subhalo mass increases, the UNet prediction improves. These examples were chosen systematically by taking the first instance in the test set with a subhalo of a particular mass bin. The first image is dimmer than the rest due to random differences in sources, unrelated to the lack of a subhalo.}
    \label{fig:examples}
\end{figure*}

\begin{figure*}
    \centering
    \includegraphics[width=.33\textwidth]{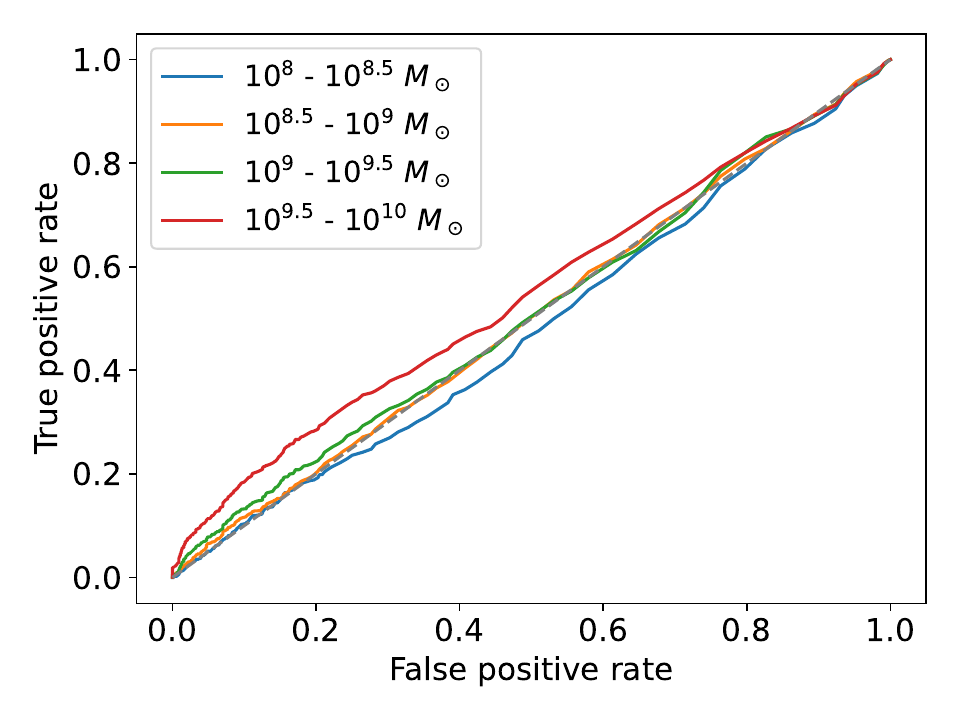}
    
    \includegraphics[width=.33\textwidth]{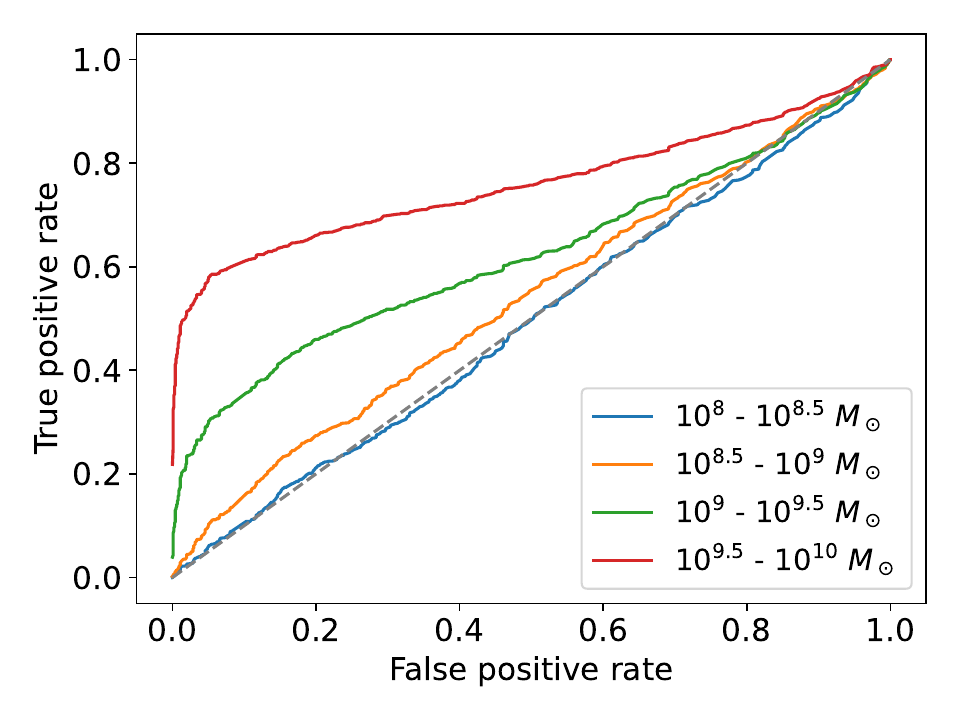}
    \includegraphics[width=.33\textwidth]{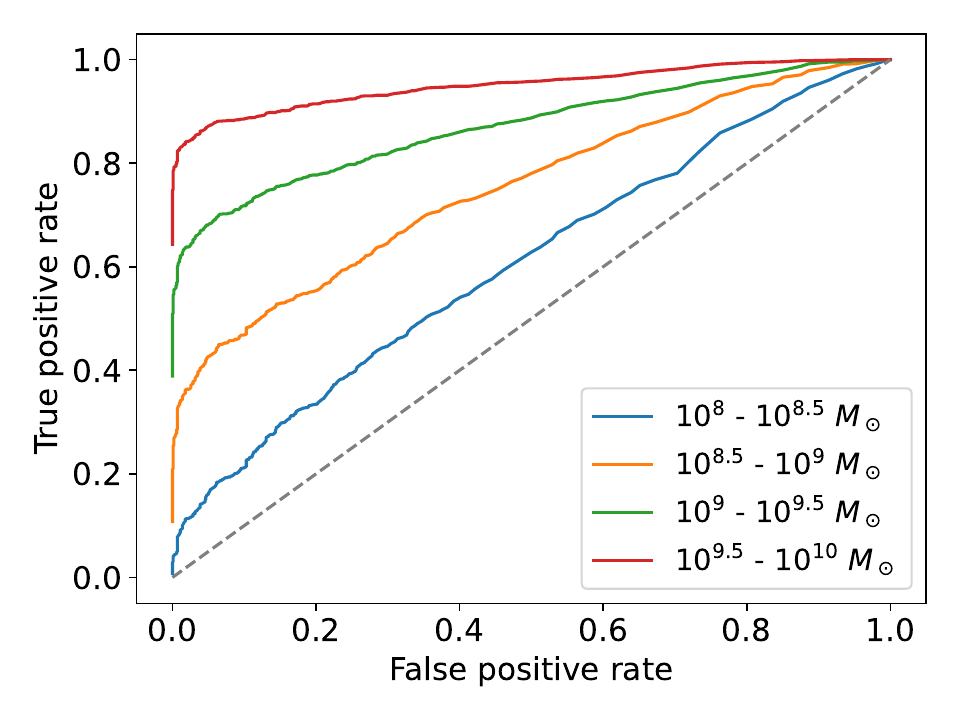}
    \includegraphics[width=.33\textwidth]{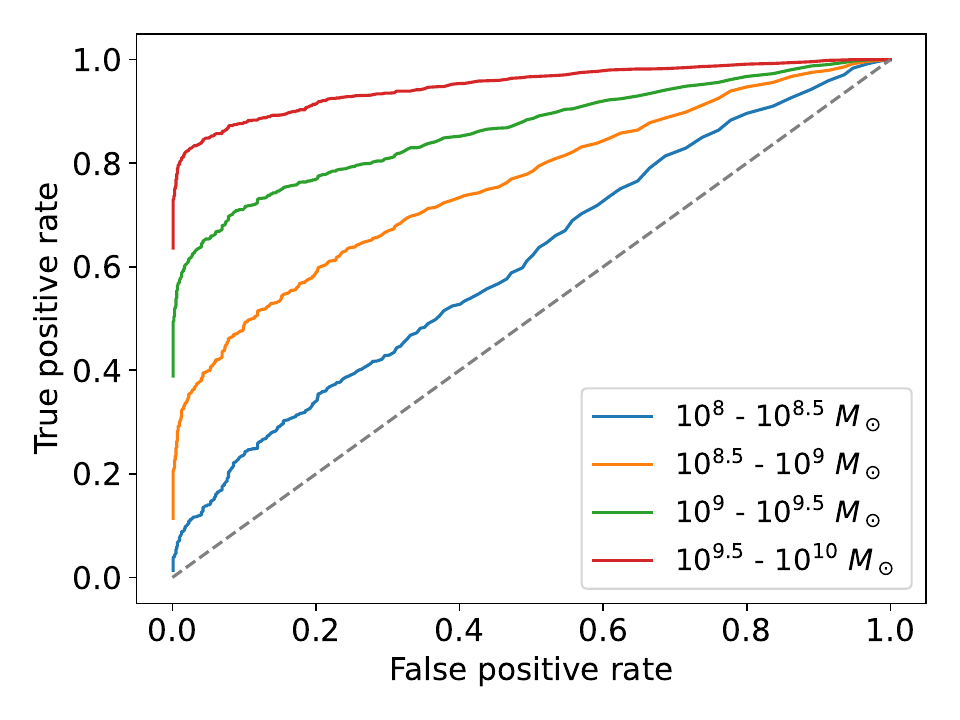}
    \caption{ROC curves at different subhalo concentrations and image resolutions. The $x$-axis shows the false positive rate, the fraction of \emph{no-subhalo} images incorrectly classified, and the $y$-axis shows the true positive rate, the fraction of \emph{subhalo} images correctly classified. Each colored curve represents a different subhalo mass bin. \emph{Top:} Low concentration ($c=15$) at high resolution ($10\un{mas}$). \emph{Bottom:} High concentration ($c = 60$) increasing in resolution from left to right ($80$, $20$, and $10\un{mas}$).}
    \label{fig:roc}
\end{figure*}

\begin{table}
    \centering
    \begin{tabular}{c c c c c}
        \toprule
         Conc. & Resolution & $10^8\text{-}10^{8.5}$ & $10^{8.5}\text{-}10^9$ & $10^9\text{-}10^{9.5}$ \\
        \midrule
         60 & 80 mas & 0.090 & 0.135 & 0.336 \\
         60 & 20 mas & 0.211 & 0.468 & 0.718 \\
         60 & 10 mas & 0.238 & 0.489 & 0.713 \\
        \midrule
         15 & 10 mas & 0.103 & 0.116 & 0.133 \\
        \bottomrule
    \end{tabular}
    \caption{The numbers on the right three columns are the true positive rates when the false positive rate is set to $0.1$ for different concentrations (first column) and resolutions (second column), separated into subhalo mass bins of $10^8\text{-}10^{8.5} \,\msun$, $10^{8.5}\text{-}10^9 \,\msun$, and $10^9\text{-}10^{9.5} \,\msun$. Our system can find high-concentration subhalos, but for low concentrations, performance is around the $0.1$ one would expect from random guessing, which always produces the same true and false positive rates. See Figure~\ref{fig:roc} for the corresponding ROC curves.}
    \label{tab:tprates}
\end{table}

We train and test on consistent subhalo concentrations and image resolutions. We show results at a concentration of 60 for all three resolutions and at a concentration of 15 for only the highest resolution, as it turns out that we are unable to detect subhalos in the latter.

We first illustrate the UNet's predictions on typical test set images in Figure \ref{fig:examples}, at high resolution (10 mas) and high concentration (60). We see qualitatively that the more massive the subhalo, the better the UNet can detect it, which is expected since higher-mass subhalos produce stronger lensing signals.

While our UNet makes predictions on a per-pixel basis, we envision it to be most useful as a binary classifier that flags which systems are worth analyzing in further detail. Thus, we evaluate the UNet as follows: we run it on each image to assign each pixel a probability of belonging to the \emph{subhalo} class, then we take the largest probability, and we call it the score of the image. We compare this score to a threshold value which we discuss below, and we classify the whole image as belonging to the \emph{subhalo} class if and only if its score falls above the threshold.
Although this metric does not account for subhalo location, the UNet predicts this accurately as well, which can be helpful for initializing a more traditional analysis. Indeed, among the \emph{subhalo} images correctly recognized by the UNet in the high-resolution, high-concentration test set, the UNet finds the subhalo location to within 0.040 arcsec (4 pixels) in more than $90\%$ of cases (the prediction being the location of the pixel assigned the highest \emph{subhalo}-class probability).

In any binary classification task, we can define a positive (in our case, \emph{subhalo}) and negative (\emph{no-subhalo}) class, and we have two goals: the classifier should label every positive example in the dataset as positive and every negative example as negative. Our success on the first goal is measured by the true positive rate (TPR), the fraction of positive examples that are correctly classified, while our failure to meet the second goal is measured by the false positive rate (FPR), the fraction of negative examples which are incorrectly classified as positive. While we want both a high TPR and a low FPR, we have to make a compromise between labeling all images positive for a perfect TPR of 1 and all negative for a perfect FPR of 0. In a receiver operating characteristic (ROC) curve, the FPR and TPR are visualized as the $x$- and $y$-axes of a graph, and we show this tradeoff by plotting a curve of every possible (FPR, TPR) as we adjust the score threshold from labeling everything \emph{no-subhalo}, $(0,0)$, to labeling everything \emph{subhalo}, $(1,1)$. These curves are monotonic by definition and the best curves are those which come closest to the (0, 1) corner. In Figure~\ref{fig:roc}, each subplot displays a different combination of subhalo concentration and image resolution, and each curve represents one of the following four subhalo-mass bins: $10^8\text{-}10^{8.5} \,\msun$, $10^{8.5}\text{-}10^9 \,\msun$, $10^9\text{-}10^{9.5} \,\msun$, and $10^{9.5}\text{-}10^{10} \,\msun$. Note that the most massive bin may correspond to subhalos luminous enough to be observed directly, so we focus especially on the performance of the second most massive bin.

In Table~\ref{tab:tprates}, we summarize the results from Figure~\ref{fig:roc}, showing the TPR when we set a threshold that gives an FPR of $10\%$. In particular, we find that under the best conditions, we can detect $10^{9}$-$10^{9.5}\,\msun$ subhalos with over a 70\% TPR.

As expected, these results show quantitatively that the UNet model is much better at detecting more massive subhalos. Furthermore, we see that the 20 mas resolution performs significantly better than 80 mas (although we see little if any improvement going from 20 to 10 mas). This suggests that higher resolution telescopes, like the future Extremely Large Telescope, can aid greatly in these searches. And finally, we see that a high concentration is crucial to detectability, in line with the findings of \citet{minor21detectability}. This selection effect may help explain why real-data detections so far have been higher than $c \sim 15$ \citep{caganjvas, zhang2024hst}. We contrast our new findings with the earlier UNet work by \citet{ostdiek22apj}, which was able to detect subhalos with a concentration of 15 in mock images, but only when the source was simple and analytical.

\begin{figure*}
    \centering
    \includegraphics[width=.33\textwidth]{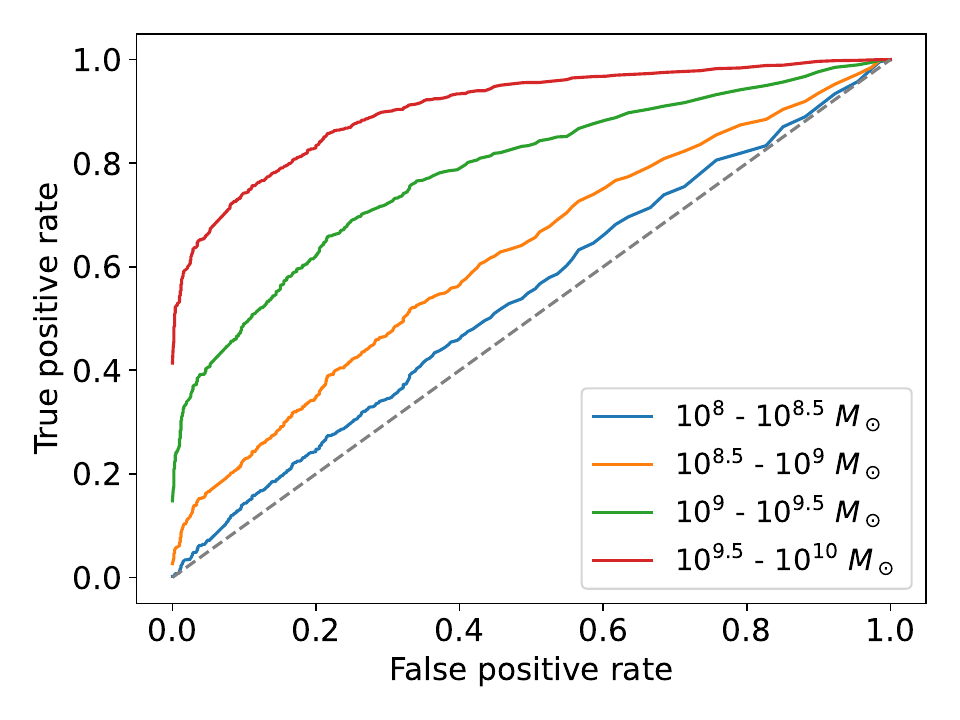}
    \includegraphics[width=.33\textwidth]{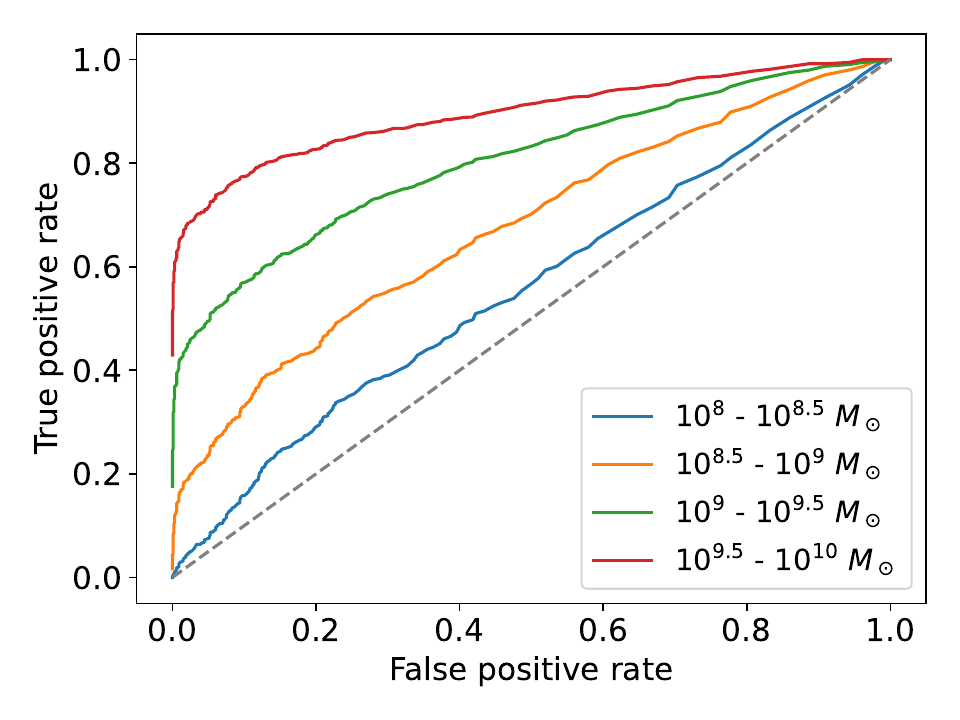}
    \includegraphics[width=.33\textwidth]{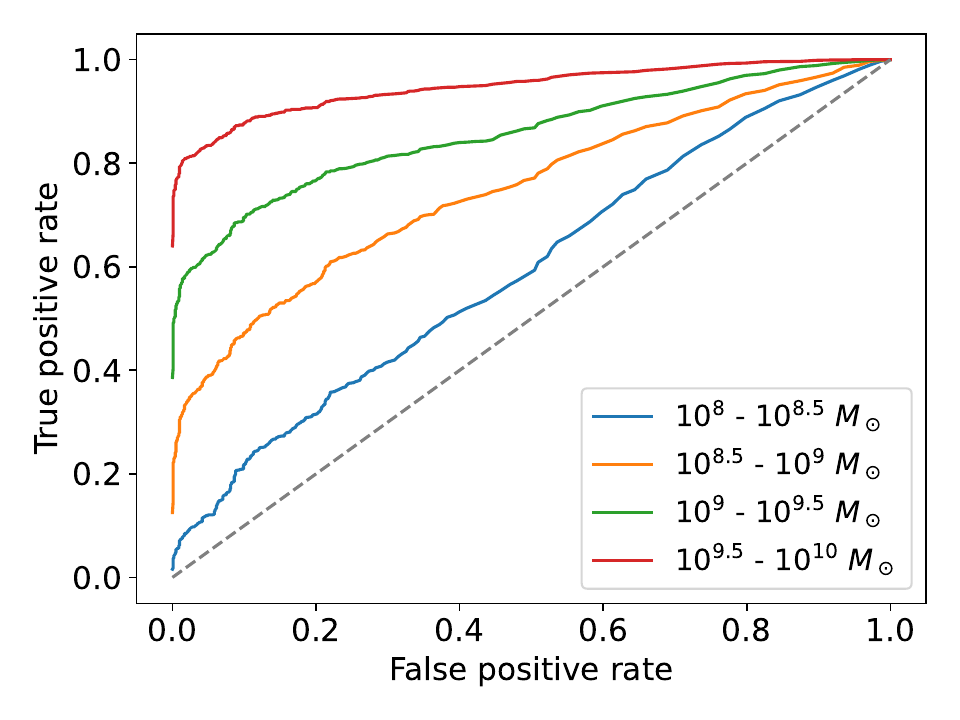}
    \caption{ROC curves of robustness tests on high concentration, high resolution data. \emph{Left:} Higher noise levels (1 orbit), test set only. \emph{Middle:} Higher noise levels (1 orbit), training and test sets. \emph{Right:} Background of 1000 low-mass subhalos between $10^6$ and $10^8\,\msun$, test set only.}
    \label{fig:robustroc}
\end{figure*}

\subsection{Robustness tests}

In addition to the above results, we run two simple robustness tests: one to increase the amount of observation noise, and the other to add a background of low-mass perturbers. Reducing simulated observation time from 10 Hubble orbits to 1, we see a significant reduction in performance: the 71.3\% true positive rate (for the $10^9$-$10^{9.5}\,\msun$ subhalo mass range) drops to 49.0\%. If we retrain on this noise level, it partially recovers to 57.0\%. This demonstrates both the importance of noise to our problem and the necessity to train the model on correct noise levels.

For the second test, we add a background of 1000 subhalos spread uniformly across the field of view, whose masses range between $10^6$ and $10^8\,\msun$, distributed as $n(m) \propto m^{\beta}$, where $m$ is the $M_{200}$ mass and $\beta = -1.9$ \citep{springel2008aquarius}. It is conceivable that even though these low-mass subhalos are not detectable individually, their combined effect could bias the predictions of a model trained without such a subhalo background. However, we find this not to be a significant issue: the 71.3\% true positive rate decreases slightly to 69.5\% once these low-mass subhalos are included. Since performance is so similar, we did not retrain on this data. The ROC curves for these robustness tests are shown in Figure~\ref{fig:robustroc}.

\section{Discussion}

As we make our mock lensing systems more realistic, we find that high concentrations become crucial for detectability. The same machine learning system which fails to detect subhalos of concentration 15, even at high mass and high resolution, can nonetheless find them again when we increase the concentration to 60.

While we have improved the source modeling and have included sufficient analytic lens complexity to produce good fits to real data, there are still ways to make our mocks more realistic. It was found that the light and mass distributions of a lens have uncorrelated ellipticities and, in certain cases, are not well-aligned \citep{shajib2019unhappy}, suggesting that the baryonic and dark matter components should be modeled separately. Furthermore, in the context of time-delay measurements of $H_0$, work by \citet{gomer2021lensmodel} suggests the macro components of true lenses ideally should be fit in a model-free way. It is still a question for future work how well our machine learning techniques generalize when lenses are taken from N-body or hydrodynamic simulations rather than generated analytically.

Furthermore, we have ignored the nuances of lens-light subtraction, which would increase both the statistical and systematic errors on our images, and we did not investigate to what extent a possible selection bias in the sampling of COSMOS galaxy sources, as well as the source and lens redshifts, would affect detectability. In particular, there can be a discrepancy if our data sample of unlensed galaxies has a different redshift distribution from the set of true source galaxies, which can be especially difficult to account for, because lensing increases sensitivity to dim, high-redshift galaxies. The COSMOS galaxy sources also contain low levels of observational noise, which we unphysically lens, but which ideally should be cleaned first while maintaining the high-resolution details of the galaxies. We estimate this to be roughly a factor of 3 times smaller than the simulated noise (corresponding to 10 orbits) added after lensing. We leave taking into account these corrections for future work.

As an idea for future exploration, machine learning with image segmentation could in principle be extended to the multiple images of a particular galaxy in a cluster lens, as has recently been done with non-machine learning methods \citep{cagan2023cab}. For example, images could be simulated using the curved arc formalism \citep{birrer2021cab}, and inputs could be fed into a UNet that takes each image as input in a separate channel.

Alternative fast techniques can also be further explored. We have tried a more physics-inspired model where we encouraged the machine learning algorithm to focus on predicting lens parameters and optimizing mass perturbations while hard coding the analytic form of the lens and the lensing equations, but we found that this usually performed worse than the direct UNet. One possible drawback of our approach is that we modeled the source galaxy by fitting shapelets rather than by using a realistic galaxy prior. It would be interesting to see if the UNet performance can be matched or superseded once generative galaxy priors \citep{lanusse2021generative,holzschuh2022generative, adam2022rim,adam2023rim} are taken into account.

In the larger context, we see machine learning as useful for flagging systems as a first pass, to narrow down the number of lenses necessary to analyze. Then, having fairly reliable priors on the location of the suspected subhalo, we would follow up with a more traditional fitting procedure in which the uncertainties are better understood. In particular, this fitting procedure would include a measurement of the effective subhalo density slope, a parameter which can provide insight into different dark matter models \citep{cagan2022slope}.
Thus, the computational advantages of machine learning, in combination with other techniques, should enable us to extract the wealth of information relevant to the small-scale nature of dark matter.

\section{Conclusions}

We found that it is possible to train a UNet, a neural network image segmentation architecture, to detect small mass perturbations (roughly $10^{8.5}\text{-}10^{9.5} \msun$)
in a galaxy-galaxy strong gravitational lens, on simulated systems with real galaxies as sources. Once trained, the UNet produces quick predictions that can scale effortlessly to the large datasets of the upcoming years. The range of possible individual subhalo in which detection using gravitational lensing can be useful is fairly narrow: we need a large enough mass to detect a perturbation, but still one small enough that the subhalo would not be directly visible. We found that our UNet can only reliably detect such subhalos at concentrations higher than those typical of $\Lambda$CDM simulations. Interestingly, these high concentrations are consistent with recent observations.

Despite the several successful sampling-based detections, this is a difficult measurement, and making sense of the upcoming data will most likely require at least two steps. First, we would use machine learning to scan over all galaxy-galaxy lenses for signs of a detectable, dark perturber. Second, for each of the systems that the first pass found promising, we would run a more robust, slower analysis, ideally free of analytic lens assumptions, to reveal reliable information about the perturber and, in particular, its mass and profile slope. It will be possible to check for any machine learning biases by running the machine learning algorithm alongside sampling-based techniques on a small random subset of the data. Finally, of course, remains the interpretation of these measurements as constraints on any particular dark matter model.

\section{Acknowledgements}
We thank Simon Birrer, Bryan Ostdiek, and Sebastian Wagner-Carena for insightful discussions.
This work was supported by the National Science Foundation under Cooperative Agreement PHY-2019786 (The NSF AI Institute for Artificial Intelligence and Fundamental Interactions, \url{http://iaifi.org/}).
The computations in this paper were run on the FASRC Cannon cluster supported by the FAS Division of Science Research Computing Group at Harvard University.

\bibliographystyle{aa}
\bibliography{main.bib}

\end{document}